# Fundamental constants explain sub-millimeter liquid shear forces

Laurence Noirez

Laboratoire Léon Brillouin (CEA-CNRS), Université Paris-Saclay, CEA-Saclay, 91191 Gif-sur-Yvette Cédex, France

Email: laurence.noirez@cea.fr

**From its links to fundamental physical constants to the identification of liquid shear elasticity, over the past decade, new ways are emerging to understand viscosity.**

This article is an invitation to no longer look at liquids in the same way. A new approach to understanding fluid flow is emerging. An unexpected link between minimum viscosity and universal constants shows that this formless matter called liquids is a daily reminder of our deep cosmic origins [1]. This fundamental link also highlights the cohesive nature of liquids, in agreement with the energy associated with the shear elastic forces recently identified at the mesoscopic scale [2], and therefore the possibility of collective effects.

The term "viscosity" is used on a daily basis. Viscosity characterizes the ability to flow. For example, knowing viscosity is vital when it comes to physiological fluids and is commonly used as an indicator to understand and treat diseases. But let's examine how it has been defined and the assumptions that underlie it.

## VISCOSITY AS AN EVOLVING CONCEPT

The concept of viscosity today has neither always been evident nor intuitive. In the mid-nineteenth century French engineer, Henri Navier, introduced the idea of viscosity as a constant representing the resistance of a layer to a wall [3]. Navier also noticed that at small scales, this resistance of a fluid to flow, depends on the substrate. He noted, for example, that water was flowing three or four times more slowly near glass than near copper. On a larger scale, the agreement between experimental results and theory gave credence to the hypothesis that the liquid has zero velocity at the substrate. The no-slip boundary assumption was adopted. Therefore, the concept of macroscopic viscosity evolved, breaking free from experimental conditions although on a small scale, it is no longer valid (for micro and nanofluidics for example) and reminds us that the transfer mechanism of interfacial stresses is still largely unknown. The link between viscosity and the strength of the molecular interactions is not straightforwardly demonstrated. This might explain the difficulty of theoreticians to develop a precise law for the interaction of liquids. The challenge is thus to find a starting point, a common characteristic, that is free from liquid molecular parameters that are highly variable in liquids and make a general theoretical framework difficult to develop. Kostya Trachenko has succeeded in identifying a nearly constant value for viscosity of various liquids at the nexus of liquid and gas states. This viscosity is independent of the molecular interactions and structure [1]. Regardless of a liquid's nature, this common characteristic--that the lowest viscosity is obtained by varying the temperature and pressure--is not a coincidence but an indication of the fundamental parameters that define the lower allowed viscosity. This enables Trachenko to define a fundamental kinematic viscosity that depends only on three fundamental physical constants: $\hbar$, $m_e$ and $m_p$ where $\hbar$ is Planck's constant, $m_e$, the electron mass and $m_p$ the proton mass [1,9]. Even separated by about 15 decades, the lower kinematic viscosity points to its fundamental origin in the more stable components of the universe.

Why does the relationship to the fundamental constants seem valid for the lowest viscosity only? Recent experimental advances might provide an answer by demonstrating that when the boundary condition is ignored conventional measurements only partially access liquid properties.

## WHAT SMALL-SCALE MEASUREMENTS TELL US

Recent measurements show that at the submillimeter scale, viscosity is not universal but depends on how a liquid interacts with a substrate [2, 4]. This result echoes the Navier observation about water's behavior on glass and copper [3]. More precisely, conventional viscosity or viscoelasticity measurements do not guarantee the necessary no-slip condition of the liquid at the wall.

When liquid/substrate wetting improves towards non-slip boundary conditions, the mesoscopic fluid response reveals a fundamentally different response [2, 4]. The fluid resists before flowing. Liquids are thus endowed with "static" shear elasticity. Why is this important? The ability of liquids to support solid-like shear stress is an indicator of their cohesion energy; Liquid shear elasticity challenges the very definition of what a liquid should be. Shear elasticity is a collective property. It involves long-range interactions (Fig. 1). Taking into account the mesoscopic shear elasticity allows a new understanding of liquids and to discover new properties on basis of collective excitations mostly applied only in solids until recently.

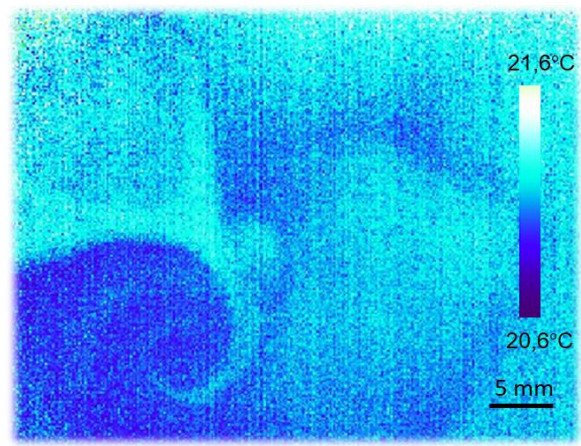

Figure 1: Liquid properties such as viscosity and shear elasticity echo the history of our distant stellar origin via their dependence on high-energy particles (1,9). In the image, the thermal traces left in the furrow of a moving object in water illustrate long-range interactive forces existing in liquids (2,4,8). Credit: P. Baroni.

## THEORY AND EXPERIMENT CONVERGE

The first experimental identification of liquid shear elasticity at the micron scale occurred in the late 1980s [5]. The research was extended to the sub-millimeter scale about 15 years later with the identification of liquid shear elasticity [2].

Concurrent to, but independent of, the work to identify liquid shear elasticity, Trachenko was applying insights from the Maxwell-Frenkel viscoelastic theory to show that nothing prevents liquids from propagating a gapped shear wave [6,7]. Zaccone and Trachenko have now developed an atomic-level mathematical theory of the (visco)elasticity of confined liquids clarifying that confinement increases liquid shear elasticity by suppressing long-wavelength shear waves [7].

Theoretical formalism and experimental measurements independently converged and established the existence of liquid shear elasticity.

## LIQUIDS NOT SO DIFFERENT FROM SOLIDS

Liquid shear elasticity is accessible at the sub-millimeter scale and under low perturbative mechanical conditions. Why? Moving away from the nearly equilibrium conditions (e.g., under flow, or by increasing the probed lengthscale or the magnitude of the shear excitation), makes its measurement hardly accessible and restores conventional viscous behavior. Should we believe that the elasticity collapses over a stress threshold? Again, our perception is wrong since liquid may not completely dissipate the mechanical energy. What becomes of it then? If energy is stored in a liquid, does it appear as a thermodynamic signature?

The first micro-thermographic images of liquids such as glycerol, polymer melts or water revealed that the liquids emit a modulated thermal signal when subjected to a mechanical shear stress wave [4,8]. Thermal waves propagate and split the liquid in hot and cold bands several tenths of a micron wide, varying synchronously with the applied stress wave. The mechanically induced temperatures are fundamentally linked to the energy of the intermolecular distances. It is a collective effect analogous to the so-called thermo-elasticity of solids, i.e., requiring strong cohesive forces.

What makes these elastic correlations so solid that thermoelastic behavior is identifiable in the so-called viscous regime? This thermomechanical coupling demonstrates that liquid molecules are not dynamically free as was often assumed at the beginning of theoretical approaches to liquids. Instead, these molecules bind as in solids, with this cohesive energy expressible through fundamental physical constants. The insight that liquids are more similar to solids than to gases, even though liquids flow and solids do not, turns out to be the key to understanding the properties of liquids. With the link between liquid properties and fundamental constants established, we can start asking what would happen to liquids and liquid-based life if the constants are given different values [1,9]. To enable liquid-based life, these constants need to lie in a constrained bio-friendly window.

This fundamental research will enable new applications in disciplines such as controlled fluid transport, thermal exchangers, health and biomedicine. Revising the definition of liquids is indispensable to extend our ability to control this state of matter. In practice, medicine is already taking advantage of this promise as it is now possible to also access the elasticity of blood [10], a "remembrance" of our distant past where only elementary particles existed [1,9].